\documentclass[final,5p,times,twocolumn]{elsarticle}
\usepackage{color}
\usepackage{amsmath}
\usepackage{amssymb}
\usepackage{subfig}
\usepackage{graphicx}
\usepackage[colorlinks=true,linkcolor=black, citecolor=blue, urlcolor=blue]{hyperref}
\usepackage{lineno,hyperref}
\biboptions{numbers,sort&compress}   % Agrupa las referencias continuas en el texto
\modulolinenumbers[5]

\begin{document}

\begin{frontmatter}

\title{Writing and storing information in an array of magnetic vortex nanodisks using their azimuthal modes}

\author{H. Vigo-Cotrina}
\address{Centro Brasileiro de Pesquisas F\'{\i}sicas, 22290-180,  Rio de Janeiro, RJ, Brazil}

\author{A.P. Guimar\~aes}
\address{Centro Brasileiro de Pesquisas F\'{\i}sicas, 22290-180,  Rio de Janeiro, RJ, Brazil}

%%\author[mymainaddress,mysecondaryaddress]{Elsevier Inc}
%%\ead[url]{www.elsevier.com}

%%\author[mysecondaryaddress]{Global Customer Service\corref{mycorrespondingauthor}}
%%\cortext[mycorrespondingauthor]{Corresponding author}
%%\ead{support@elsevier.com}

%%\address[mymainaddress]{1600 John F Kennedy Boulevard, Philadelphia}
%%\address[mysecondaryaddress]{360 Park Avenue South, New York}

\begin{abstract}
The switching of a vortex core of a single disk in an array of a multilayer system is investigated by micromagnetic simulation. We found that the perpendicular uniaxial anisotropy (PUA) decreases the frequencies of the azimuthal mode in disks with magnetic vortex configuration. We obtained a phase diagram of magnetic field intensity vs. frequency of the azimuthal mode, as a function of the value of perpendicular uniaxial anisotropy. We demonstrated that rotating magnetic fields (CW and CCW) with frequency equal to azimuthal modes can be used to switch the vortex core of single disks in a disk array. This allows obtaining different memory states with a single array of nanodisks, and therefore writing information through the application of rotating fields.

\end{abstract}


\end{frontmatter}

%\linenumbers

\section{Introduction}\label{introduction}
\indent The magnetic vortex configuration is characterized by an in plane curling magnetization,  and a core, where the magnetization points out of the plane. The curling direction defines the circulation C = +1 (counterclockwise (CCW)) and C = -1 (clockwise (CW)). The core has polarity p = +1 when it points along the +z direction and p = -1 in the -z direction \cite{Guslienko:2008}.\\
\indent A magnetic vortex presents a translation mode of low frequency, in the sub-gigahertz range, known as gyrotropic motion \cite{Guslienko:2008,Guimaraes:2009} and other two modes of higher frequency ($>$ 1$\,$GHz): azimuthal and radial modes, which have their origin in magnetostatic interactions and thus are also dependent on the dimensions of the disk \cite{Guslienko2008(2),Awad2010,Markus2014}.\\
\indent Depending on the ratio of the thickness to the radius of the disk ($\beta$ = L/R), the azimuthal modes have a splitting in the frequencies \cite{Park2005}. These frequencies have CCW and CW senses of rotation \cite{Guslienko2008(2)}.\\
\indent Magnetic vortices have many potential applications in magnetic data storage devices \cite{Bohlens2008,Guimaraes:2009,Jung2012,Helmunt2017}. For example, a vortex with p = +1 can store bit 1,  and a vortex with p = -1 can store bit 0, or vice-versa. In these applications, the issue of switching vortex cores is a topic of great interest, that has been studied for a long time \cite{Yoo2010,Kammerer2011,Yoo2012,Kammerer2012,Woo:2015,Fior:2016}.\\
\indent Using rotating magnetic fields with a frequency equal to the gyrotropic frequency, it is possible to switch the vortex core polarity \cite{Fior:2016,Lee2008}, but with the downside that this only happens when the sense of rotation of the gyrotropic motion (which is determined by p) coincides with the sense of rotation of the magnetic field \cite{Yoo2010,Fior:2016,Noske2014}. Another proposal found in the literature is to use rotating magnetic fields with frequencies equal to the characteristic frequency of the azimuthal modes \cite{Kammerer2011,Woo:2015}. This method has the great advantage that magnetic rotating fields can be used with both directions of rotation (CCW and CW), and allow shorter switching times \cite{Woo:2015,Kammerer2011}.\\
\indent Switching of vortex cores using azimuthal modes allows high switching critical velocities (of the order of $\sim$ 800$\,$m/s) \cite{Kammerer2011}, compared to those of the gyrotropic mode ($\sim$ 330$\,$m/s) \cite{Lee2008}.\\ 
\indent Nanodisks with a magnetic vortex configuration are generally produced by nanolithography in the form of arrays (matrices of nanodisks) on substrates that can influence their dynamic properties. For example, in a multilayer system, a perpendicular uniaxial anisotropy (PUA) can be induced due to the interface contribution, as has already been demonstrated by Garcia \textit{et al.} \cite{Garcia:2010}. This PUA influences the processes of switching of the vortex core \cite{Novais:2013,Fior:2016}. \\
\indent An array of disks can be used to build an information storage device and/or build logic gate circuits  \cite{Jung2012,Helmunt2017}. In these arrays, the polarity has an important role, since it determines the type of logical gate to be obtained \cite{Jung2012}. Consequently, it is necessary to search for mechanisms to control the polarity of a single disk in an array, without altering the polarity of the neighboring disks.\\
\indent The goal of this work is to propose a novel method for controlling the selectively switching of one single vortex core in a matrix of  nanodisk multilayer system, in order to obtain the desired combinations of bits in this matrix. For this purpose, we have used micromagnetic simulation. All simulations were made using the open source software Mumax3 \cite{Vansteenkiste:2014}, with discretization cell size of 2 $\times$ 2 $\times$ L nm$^3$, where L is the thickness of the disk. The material used is Permalloy (NiFe), with typical parameters \cite{Guslienko:2002, Guslienko:2008, Novais:2013}: saturation magnetization M$_s$ = 8.6 $\times$ 10$^5$ A/m$^2$, exchange stiffness $A$ = 1.3 $\times$ 10$^{-11}$ J/m, and damping constant $\alpha$ = 0.01. The perpendicular uniaxial anisotropy constant\footnote{These values of K$_z$ can be obtained experimentally increasing the thickness of disk as shown in ref. \cite{Garcia:2010}.} (K$_z$) varied from 0 to 200$\,$kJ/m$^3$.  For larger values of K$_z$, skyrmion type magnetic structures  emerge \cite{Fior:2016,Novais:2013}.

\section{Results and discussion}
\subsection{Isolated disk} 
\begin{figure}[h]
\centering
\includegraphics[width=1\columnwidth]{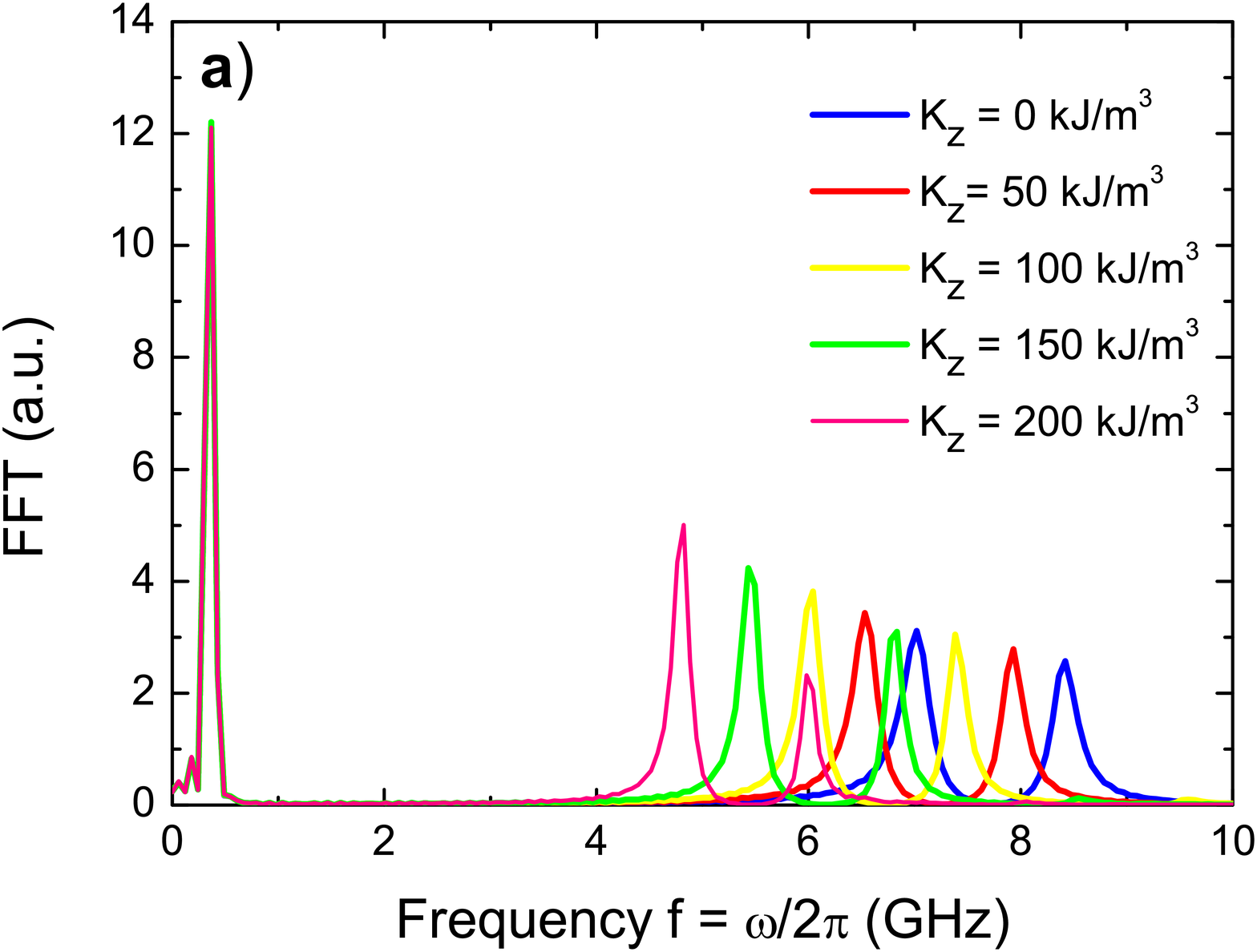}
\includegraphics[width=1\columnwidth]{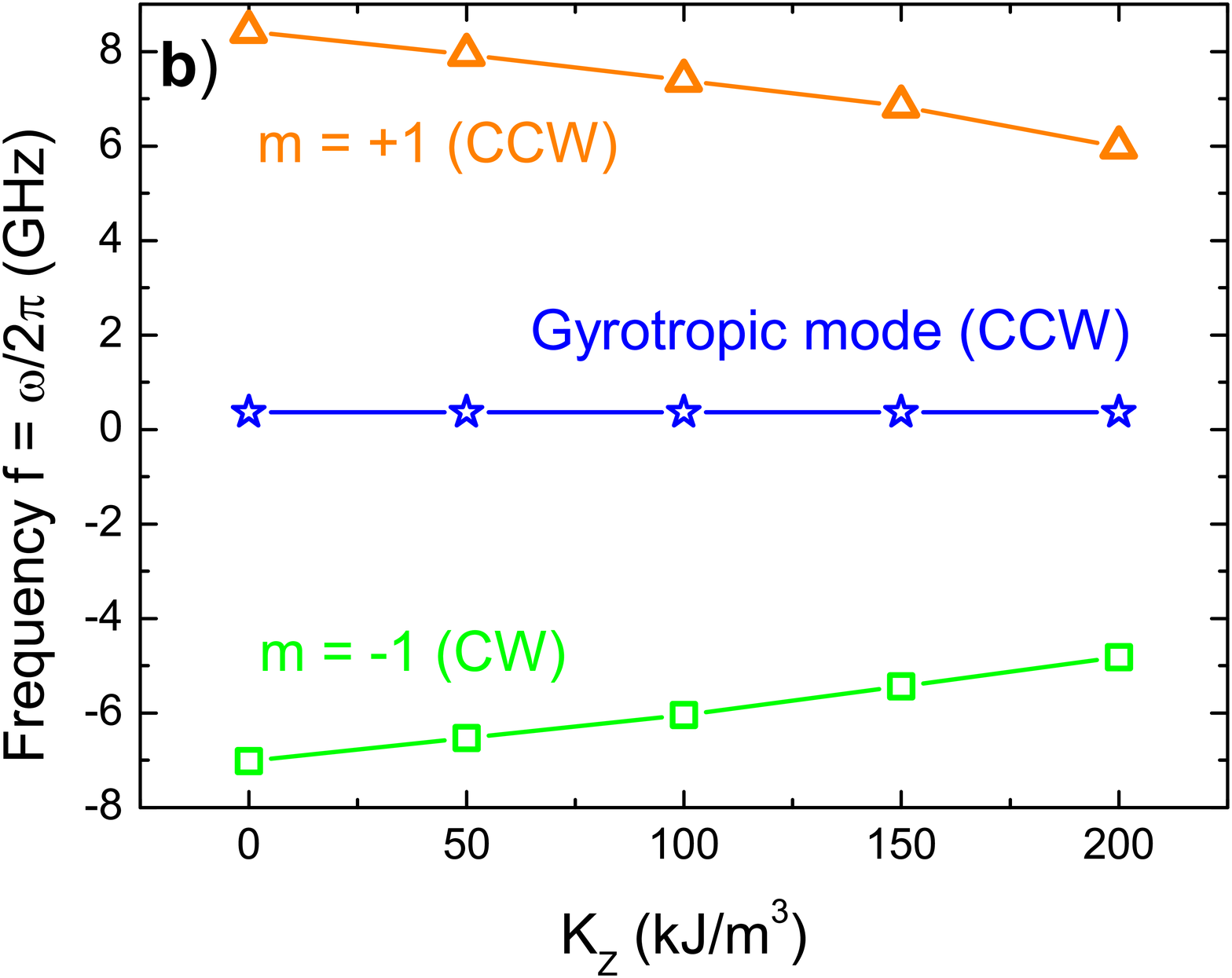}
\caption{(a) and (b) Values of the gyrotropic mode frequency and azimuthal spin wave frequencies for m = +1 (counterclockwise) and m = -1 (clockwise) obtained by a fast Fourier transform (FFT) from the time evolution of the x-component of the magnetization for each value of K$_z$, for p = +1.  Orange triangles correspond to the counterclockwise frequencies, green squares to the clockwise frequencies, and blue stars to the gyrotropic mode.}\label{fft}
\end{figure}

We used disks with thickness L = 20$\,$nm  and diameter D = 500$\,$nm. For these dimensions, the magnetic vortex configuration is stable \cite{Fior:2016}. We assumed that the vortex core is initially at the equilibrium position at the center of the disk, and has polarity p = +1 and circulation c = +1.\\
\indent In order to excite the azimuthal spin wave modes, we have applied an in-plane sinc pulse magnetic field \textbf{B}(t) = (B$_0\sin$(x)/x,0,0), with x = 2$\pi$f(t-t$_0$), centered on t$_0$ = 1$\,$ns, where B$_0$ = 1$\,$mT is the magnetic field amplitude and f = 50$\,$GHz is the frequency of the magnetic field pulse. The frequencies of the modes are obtained by  fast Fourier Transform (FFT) from the time evolution of the x-component of the magnetization. These frequencies are shown in Fig. \ref{fft}. We repeat the same procedure for each value of K$_z$. 

\indent There are three frequencies for each value of K$_z$. The lowest value frequency corresponds to the gyrotropic mode (f$_0$ $\approx$ 0.35$\,$GHz) and the other two frequencies correspond to the m = -1 (clockwise) and m = +1 (counterclockwise) azimuthal modes, respectively \cite{Kammerer2011, Woo:2015}.\\
\indent The frequency of the gyrotropic mode remains almost constant with the increase\footnote{There is a variation of approximately 3$\%$ for the maximum value of K$_z$, but this change is negligible. This too was demonstrated by Fior \textit{et al.} \cite{Fior:2016}.} of K$_z$. The azimuthal frequencies decrease with increasing anisotropy, as shown in Fig. \ref{fft}(a), because the influence of PUA modifies the configuration of the magnetic vortex  \cite{Garcia:2010,Novais:2013}. It is important to note that the effect of PUA on the vortex configuration is totally different from that produced by a perpendicular magnetic field (PMF). Whereas PUA does not alter the gyrotropic mode, PMF does so in a manner proportional to the intensity  of this  field \cite{Woo:2015}.\\
\indent In Fig. \ref{fft}(b) are shown all the frequencies (f = $\omega$/2$\pi$), considering negative values for m = -1 and positive values for m = +1 \cite{Woo:2015}.\\
\indent In order to switch the vortex core, we have used an in-plane rotating magnetic field $\textbf{B}(t) = B_0\cos(\omega t)\hat{x} + B_0\sin(\omega t)\hat{y}$ ($+\omega$ for CCW and $-\omega$ for CW ) bursts with duration of 24 periods, as suggested by Kammerer \textit{et al.} \cite{Kammerer2011}. After the magnetic field is turned off, we have monitored the micromagnetic simulation for an additional 1.5$\,$ns with zero magnetic field, to observe possible switching that may occur due to delayed processes \cite{Kammerer2013}.\\
\indent We start by exploring the switching vortex core  using f = 0.35$\,$GHz (gyrotropic mode); we encountered a minimal magnetic field intensity $B_0$ = 1.2$\,$mT, to obtain the switching of vortex cores for the entire range of perpendicular uniaxial anisotropy constant (K$_z$) used in this work (see section \ref{introduction}).\\

\begin{figure}[h]
\centering
\includegraphics[width=1\columnwidth]{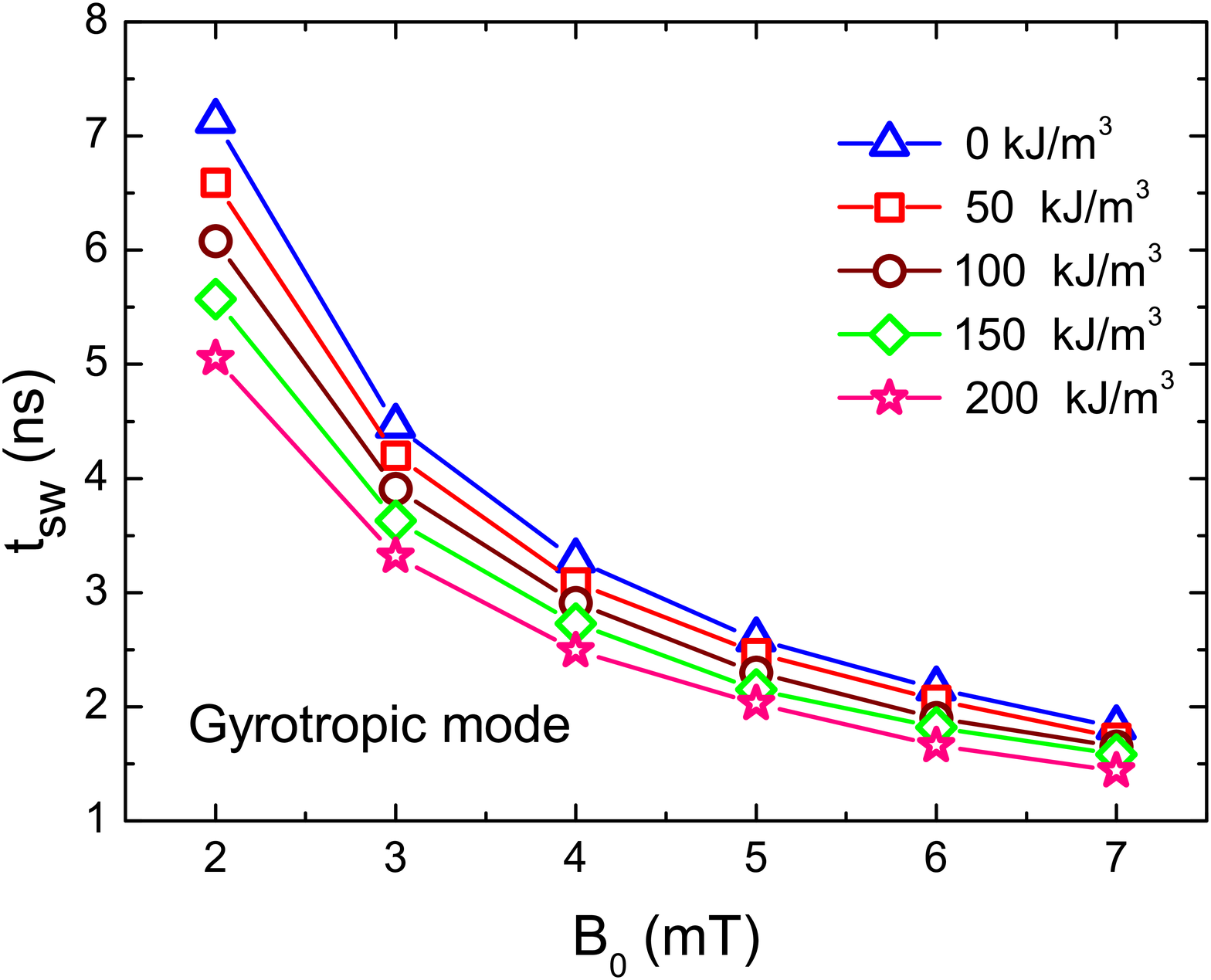}
\caption{Switching times trough the gyrotropic mode as a function of magnetic field intensity (B$_0$) for different values of K$_z$.}\label{girotropico}
\end{figure}

\begin{figure}[h]
\centering
\includegraphics[width=1\columnwidth]{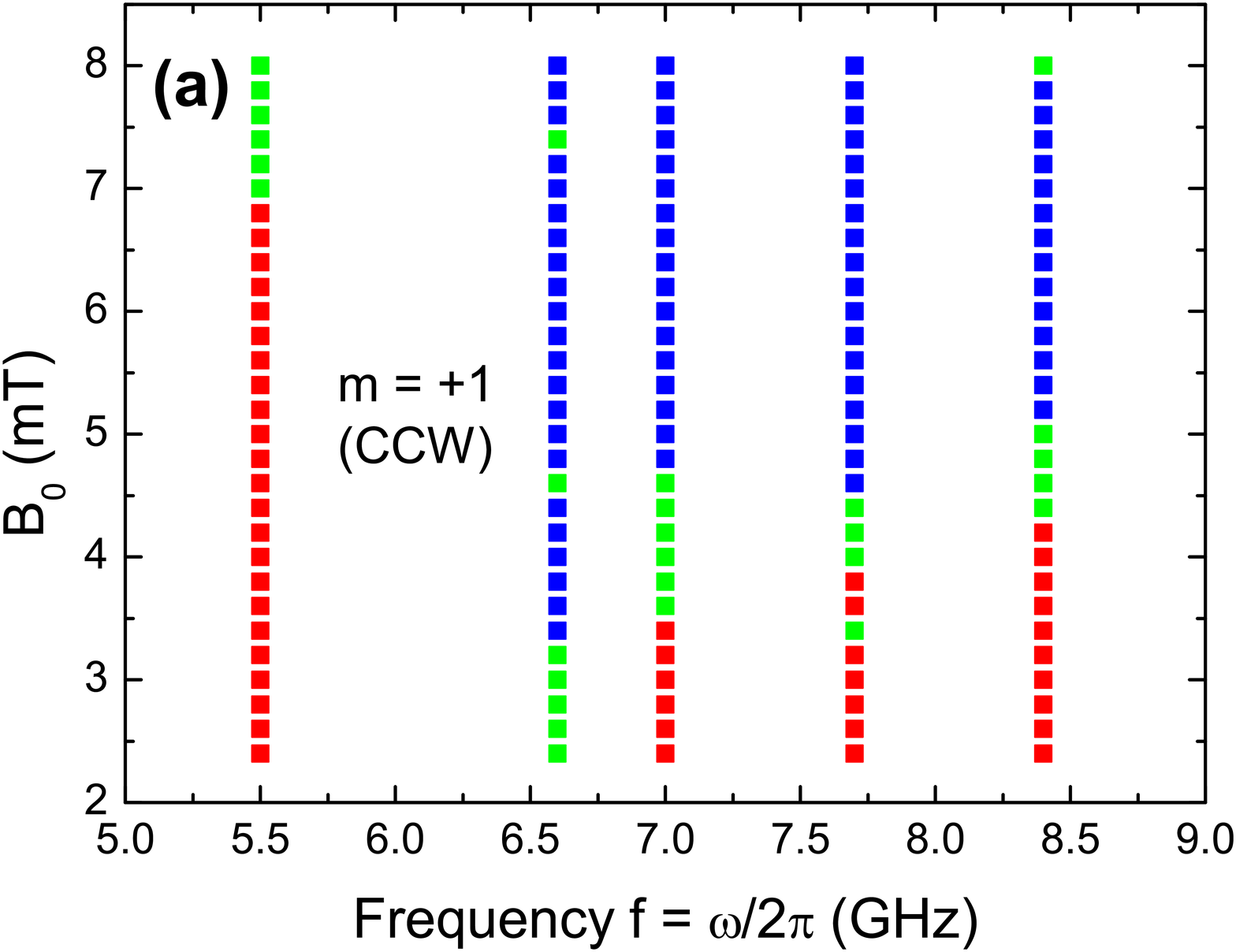}
\includegraphics[width=1\columnwidth]{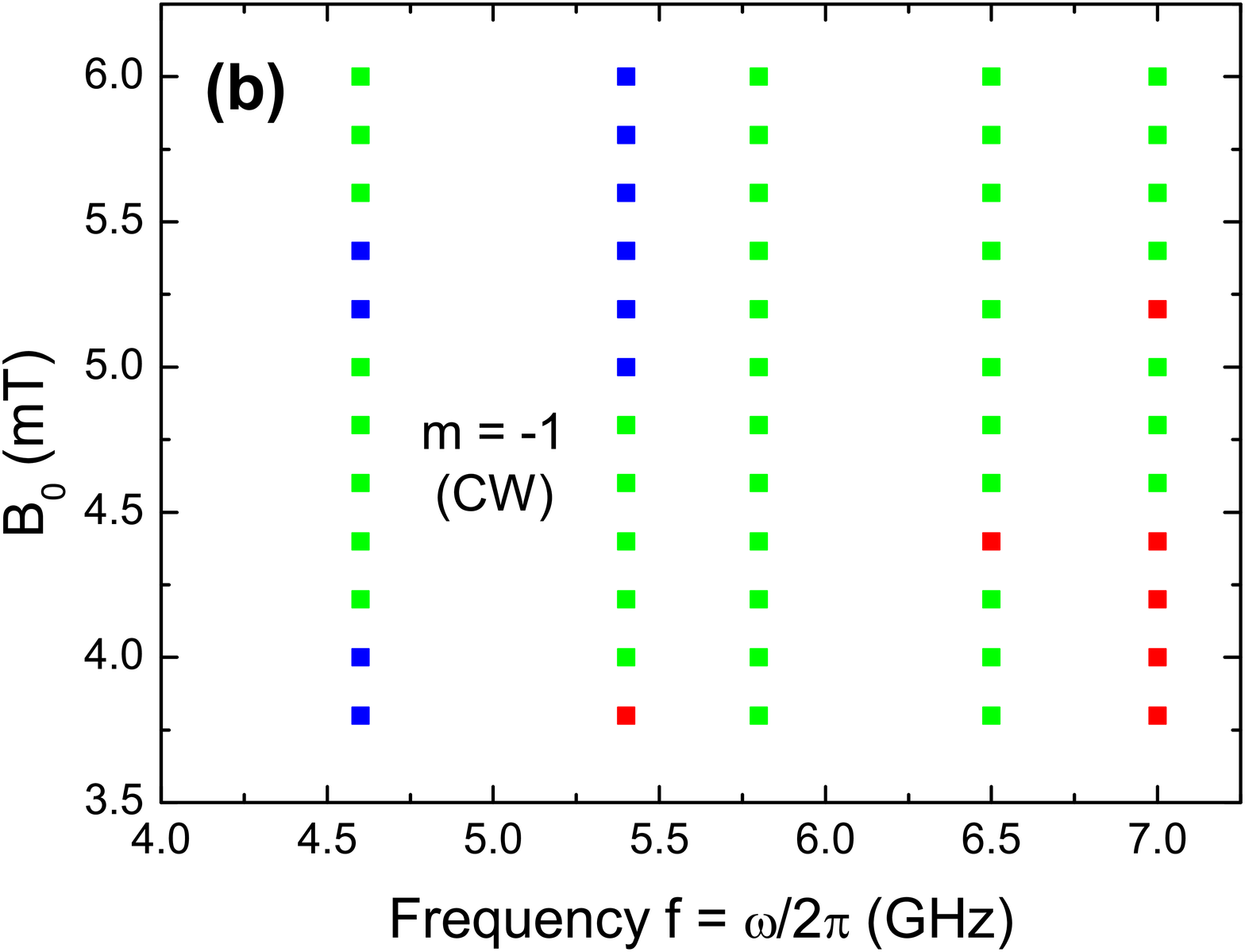}
\caption{Switching phase diagrams for (a) m = +1 (CCW) and (b) m = -1. Red triangles indicate no switching, green squares indicate single switching and blue squares indicate multiples switching. }\label{fase}
\end{figure}

\indent In Fig. \ref{girotropico} are shown the switching times\footnote{See Supplementary material for details of how these values were obtained these values.} (t$_{sw}$) as a function of B$_0$, for each value of K$_z$ used in this work, using the gyrotropic mode. We found a decrease of values of t$_{sw}$ with the increase of B$_0$.\\
\indent For K$_z$ = 0$\,$kJ/m$^3$ it is obtained a switching time (t$_{sw}$) of approximately 15$\,$ns (Fig. \ref{girotropico}). This time is reduced by approximately 88$\%$ with the increase of B$_0$ from 15$\,$ns (B$_0$ = 1.2$\,$mT) to 1.82$\,$ns (B$_0$ = 7$\,$mT). For larger intensity magnetic field, undesirable multiple switching events appear. The same behavior is observed for K$_z$ $\neq$ 0.\\
\indent Althought t$_{sw}$ decreases with the increase of B$_0$, the critical velocity (approximately 329$\,$m/s) that the vortex core reaches  before switching, is the same for all values of B$_0$. This is known as the universal criterion of switching, as has already been demonstrated by Lee \textit{et al.} \cite{Lee2008}. However, this critical velocity decreases when K$_z$ $\neq$ 0, from 329$\,$m/s (K$_z$ = 0$\,$kJ/m$^3$) to 200$\,$m/s (K$_z$ = 200$\,$kJ/m$^3$), but is still independent of B$_0$.  These values are consistent with those obtained by Fior \textit{et al.} \cite{Fior:2016}.\\
\indent In order to obtain the magnetic field intensity  to switch the vortex core using azimuthal modes, we have varied B$_0$ from 1$\,$mT to 6$\,$mT for m = -1 mode, and from 1$\,$mT to 8$\,$mT for the m = +1 mode, in 0.2$\,$mT steps for both m = +1 (CCW) and m = -1 (CW) modes. The switching phase diagram (B$_0$ vs. frequency) is shown in Fig. \ref{fase}. We have slightly varied the values of the frequencies shown in Fig. \ref{fft}, as suggested in ref. \cite{Woo:2015}, in order to obtain lower values of the threshold magnetic field intensity. \\

\begin{figure}[h]
\centering
\includegraphics[width=1\columnwidth]{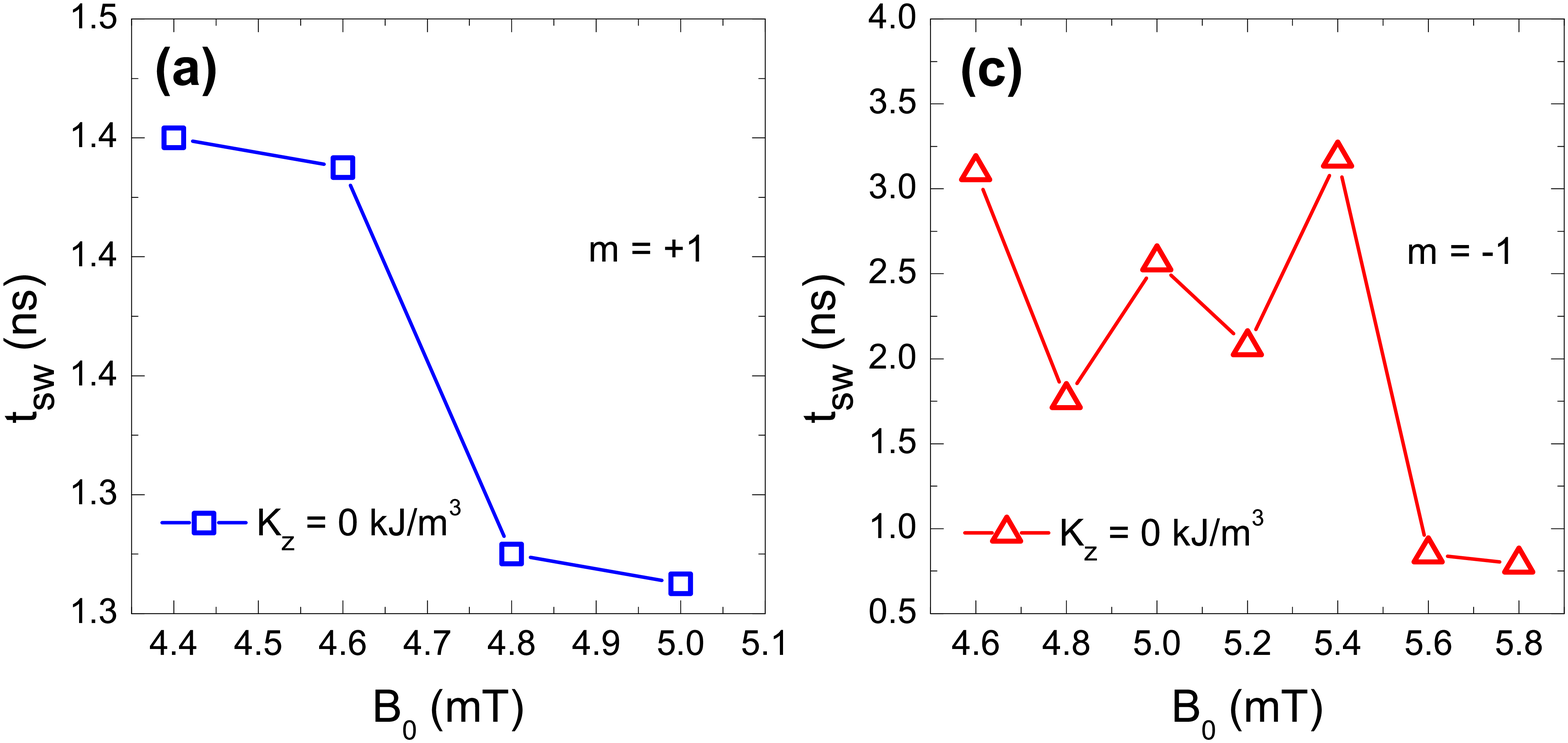}
\includegraphics[width=1\columnwidth]{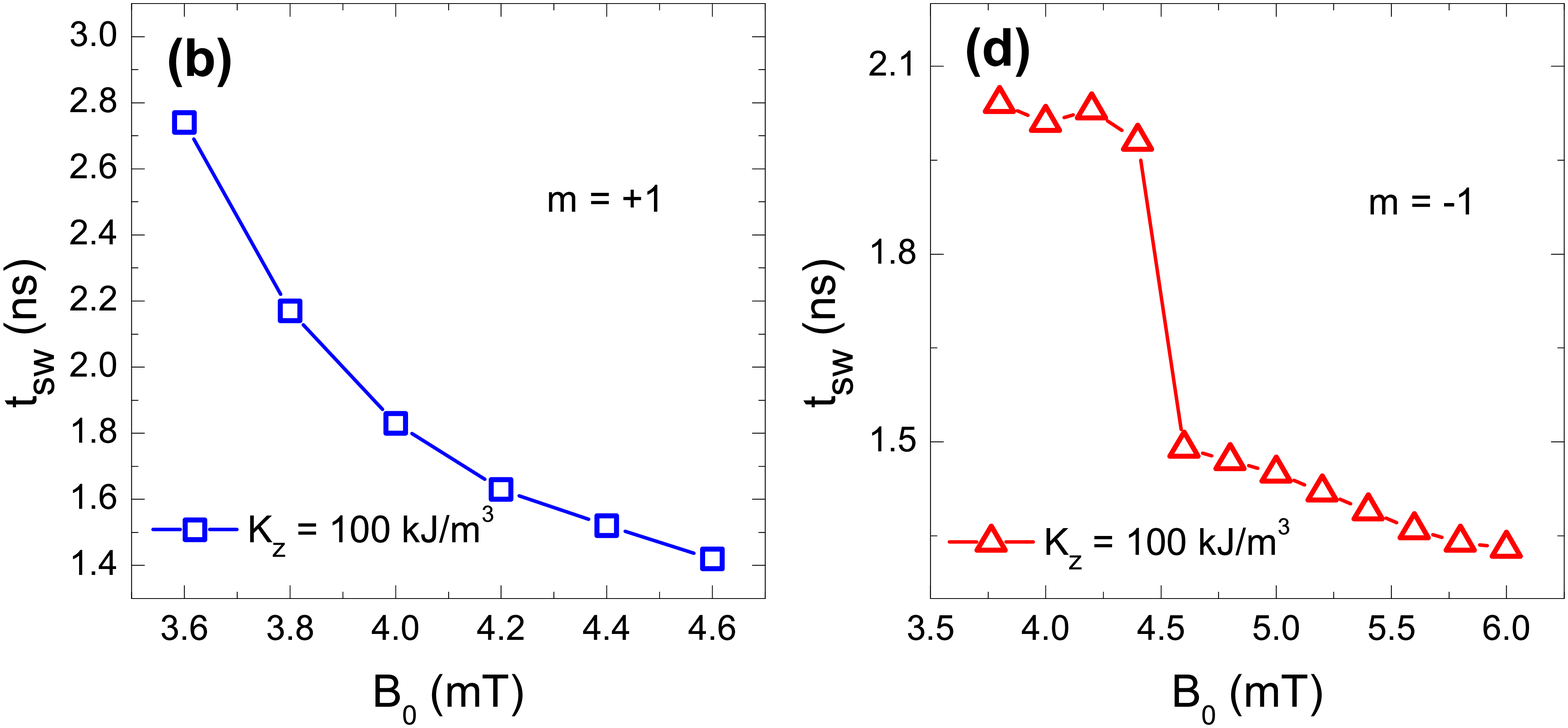}
\caption{Switching times versus magnetic field intensity for (a-b) m = +1 (CCW) and (c-d) m = -1 (CW), for K$_z$ = 0$\,$kJ/m$^3$ and K$_z$ = 100$\,$kJ/m$^3$.}\label{tiempo}
\end{figure}

\indent We encountered three regions for both modes m = +1 and m = -1: 1) no switching, 2) single switching, and 3) multiple switching. We are interested in the region of single switching, with the purpose of having  total control in the selectivity of the resulting polarity p.\\
\indent In Fig. \ref{fase} we do not show the region for 1$\,$mT $<$ B$_0$ $<$ 2$\,$mT for m = +1 mode, and the region for 1$\,$mT $<$ B$_0$ $<$ 3.6$\,$mT for m = -1 mode, since there is no switching for all the values of K$_z$ used in this work.\\
\indent The threshold of the  magnetic field  intensity (B$_0$) and the range of single switching of vortex core are different for each value of K$_z$ (see Fig. \ref{fase}), and for modes m = +1 and m = -1. 
\indent A wider range of B$_0$ values is found for m = -1 mode that result in a single switching (green squares), in comparison with m = +1. For m = +1 mode, multiple switching events are dominant in the phase diagram, whereas for m = -1, single switching events are more frequent.\\
\indent Multiple switching events appear because the applied magnetic field pumps enough energy to reverse repeatedly the vortex core between p = +1 and p = -1 \cite{Kammerer2012}.\\
\indent For the m = +1 mode, we used a magnetic field (B$_0$) of up to 8$\,$mT  to obtain  single switching for K$_z$ = 200$\,$kJ/m$^3$. This is different for the  m = -1 mode, where a threshold of magnetic field  B$_0$ = 3.6$\,$mT is necessary to obtain a single switching of vortex core.\\
\indent All these differences between modes m = +1 and m = -1 are due to the fact that the modes act differently in the creation of a dip, which is the first step in the switching process  \cite{Guslienko2008(3),Kammerer2011}. Whereas that m = -1  mode leads to the formation of a single dip, m = +1 mode leads to the formation of a double dip \cite{Kammerer2011}.\\
\indent Fig. \ref{tiempo} shows the switching times (t$_{sw}$) for both modes, and for the case K$_z$ = 0$\,$kJ/m$^3$ and K$_z$ = 100$\,$kJ/m$^3$. These times decrease with increasing intensity of the magnetic field for m = +1 mode, however, for m = -1, it is observed  that for some values of B$_0$, t$_{sw}$ does not have the same behavior\footnote{Supplemental material shows t$_{sw}$ for all values of K$_z$ used in this work.}. This can also be attributed to nolinear dynamics.\\
\indent Similar behaviors were obtained by Kammerer \textit{et al.} \cite{Kammerer2011} for Permalloy disks, and using K$_z$ = 0$\,$kJ/m$^3$.

\indent It is important to note that although  the switching time can be reduced increasing the magnetic field intensity, and using the gyrotropic mode (Fig. \ref{girotropico}), we obtain shorter switching times with lower field intensities using the azimuthal modes (Fig. \ref{tiempo}). As mentioned earlier, we get t$_{sw}$ = 1.82$\,$ns (B$_0$ = 7$\,$mT) using the gyrotropic mode for K$_z$ = 0$\,$kJ/m$^3$, but using the azimuthal mode  we get shorter times, of approximately t$_{sw}$ = 1.29$\,$ns for the m = +1 mode, and t$_{sw}$ = 0.85$\,$ns for m = -1 mode. Both values of t$_{sw}$ were obtained with  magnetic field intensity smaller that 7$\,$mT (Fig. \ref{tiempo}).\\
\indent The nolinear dynamics breaks the universal criterion for switching of vortex cores found for the  gyrotropic mode. We have obtained different values for critical velocities. For example, for K$_z$ = 0$\,$kJ/m$^3$, we found an average critical velocity (v$_{sw}$) for the entire region where there is  single switching (Fig. \ref{fase}) of approximately 816$\,$m/s and 400$\,$m/s, for modes m = +1 and m = -1,  respectively. These values are higher than those found in the gyrotropic mode, and similar to those found in ref. \cite{Kammerer2011}.\\
\indent In Table \ref{tabla1} are shown the values of the average critical velocities for modes  m = +1 and m = -1.

\begin{table}[h!]
\centering
\begin{tabular}{|c|c|c|c|} \hline

K$_z$ (kJ/m$^3$)	&  v$_{sw}$ (m/s) 	& v$_{sw}$ (m/s)\\
 & m = +1 & m = -1 \\\hline
0		&	816	& 400 \\\hline

50		&	913	& 380 \\\hline

100	&	686	&   381\\\hline

150	&	806	& 491 \\\hline

200	&	601	& 375 \\\hline
\end{tabular}
\caption{Average critical velocity for different values of K$_z$ and  modes m = +1 and m = -1.}\label{tabla1}
\end{table}

\indent All values of critical velocities shown in Table \ref{tabla1}, using the azimuthal modes, are higher than the universal critical velocity found using the gyrotropic mode, of approximately 330$\,$m/s. These higher velocities are responsible for the shorter switching times.\\
\indent The average critical velocities do not show a linear behavior  with the increase of K$_z$; they increase for K$_z$ = 50$\,$kJ/m$^3$ (m = +1),  then decrease for K$_z$ = 100$\,$kJ/m$^3$ and then increase again. This behavior contrasts with the case where PMF increases or decreases v$_{sw}$ depending on whether the PMF is parallel or antiparallel to p \cite{Woo:2015}.\\
\indent This difference is expected, since PUA does not modify the gyrotropic mode, whereas PMF does. Moreover, PUA and PMF act differently on the vortex core, leading to totally different behaviors in the switching processes, as has already been demonstrated by Fior \textit{et al.} \cite{Fior:2016}.\\
\indent Next, we used the influence of PUA and  rotating magnetic fields in order to obtain different final states in a 2$\times$2 matrix  of disks.

\subsection{2 $\times$ 2 matrix}
\begin{figure}[h]
\centering
\includegraphics[width=4cm,height =4cm]{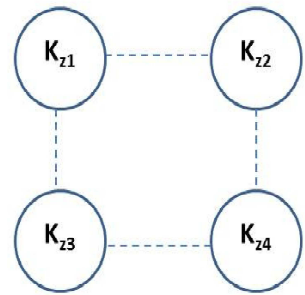}
\caption{Array of disks with magnetic vortex configuration and different values of K$_z$.}\label{discos}
\end{figure}

\indent We will now describe a matrix of four vortex disks, where we can write four bits of information using the rotating magnetic fields.\\
\indent We used an array of four identical disks, as shown in Fig. \ref{discos}, with thickness L = 20$\,$nm, diameter D = 500$\,$nm and separated by an edge to edge distance x = 500$\,$nm. Each disk has its own K$_{zn}$, with n = 1,2,3,4 and K$_{z1}$ $<$ K$_{z2}$ $<$ K$_{z3}$ $<$ K$_{z4}$. The initial configuration is that all disks have polarity and circulation p = C = +1. It is important to mention that the magnetostatic interation between the disks can change the values of the intensity of magnetic fields for switching \cite{Yao2011} shown in Fig. \ref{fase}. However, this does not alter the principle that the magnetic field having frequency  equal to either modes (m = +1 or m = -1) only reverses the vortex core of the disk to which these modes correspond.

\begin{figure}[h]
\centering
\includegraphics[width=1\columnwidth]{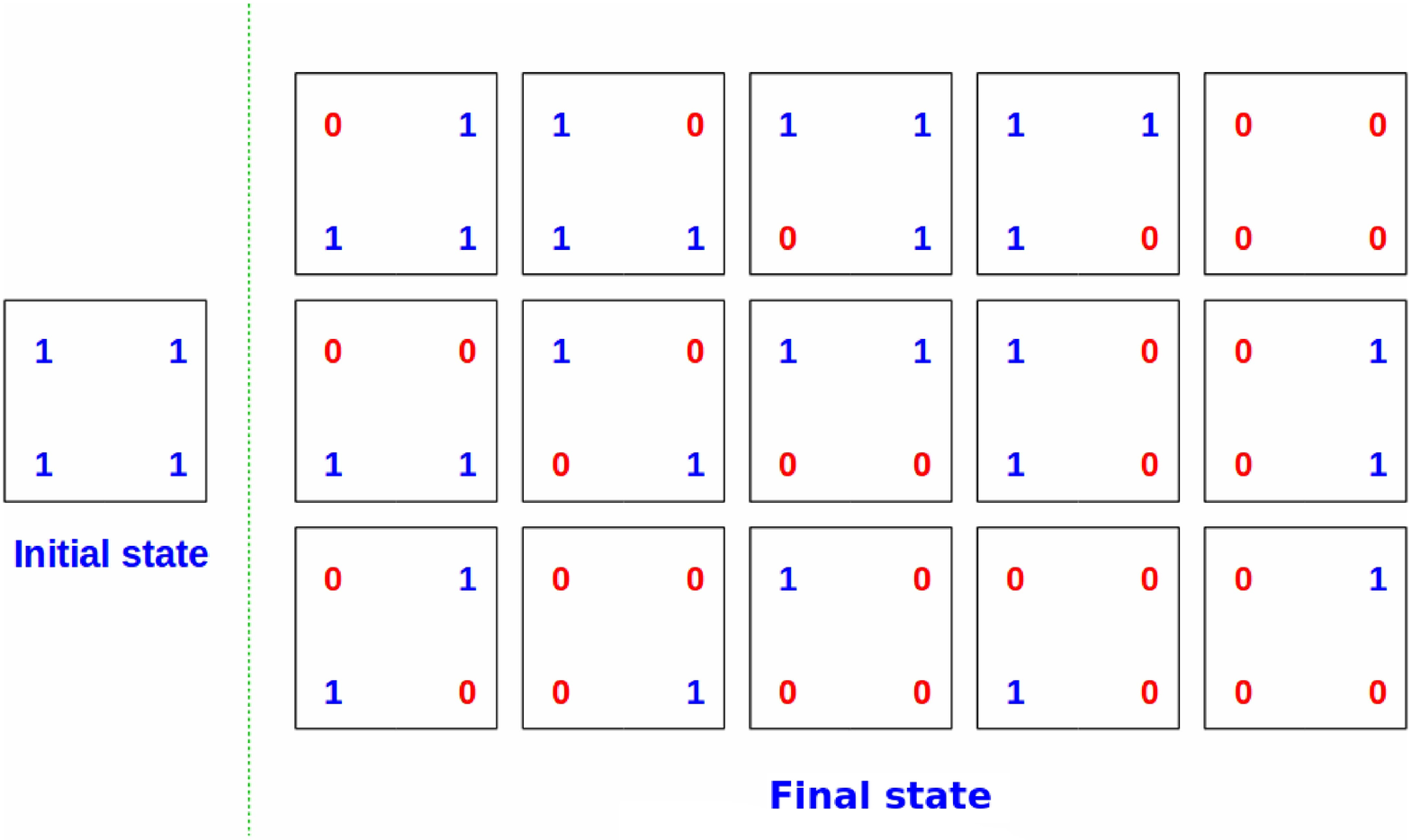}
\caption{Initial configuration of the matrix of disks and the resulting configurations obtained after applying the rotating fields. Blue number 1 and red number 0 correspond to polarity p = +1 and p = -1, respectively}\label{cartas}
\end{figure}
\indent In order to obtain different final states, as shown in Fig. \ref{cartas}, we used a global CCW rotating magnetic field $\textbf{B}(t) = B_0\cos(\omega t)\hat{x} + B_0\sin(\omega t)\hat{y}$ acting on the entire matrix, with duration of 24 periods.\\
\indent We used the following convention: blue digit 1 to indicate positive polarity p = +1, and red digit 0 to indicate negative polarity p = -1 (see Fig. \ref{cartas}).\\
\indent In order to obtain any of the final states that have a single red digit, we choose a frequency f equal to the azimuthal m = +1 mode of the disk of interest to be switched. For example, to switch the  vortex core of disk 1 (K$_{z1}$ = 0$\,$kJ/m$^3$), we used a frequency f = 8.4$\,$GHz (see Fig. \ref{fft}),  and B$_0$ = 4.6$\,$mT. The magnetic field will efficiently excite the disk 1, to which the f frequency corresponds, switching the vortex core  only in this disk (see videos in the Supplementary material). It is important to remark that this would be impossible using a frequency equal to the gyrotropic mode, because disks with higher K$_z$ would switch before those of smaller values of K$_z$ \cite{Fior:2016}.\\
\indent A final state with two red digits can be obtained in two steps using the azimuthal modes: first switching one of the disks,  and next the second disk. This could be done in one step using the gyrotropic mode, depending on which disks the user wants the switching. For example, if one desires to switch only disks 3 and disk 4, the duration of applied  magnetic field will have to be that necessary for the switching to occur on disk 3, since the switching on disk 4 would occur earlier \cite{Fior:2016}, due to K$_{z3}$ $<$ K$_{z4}$, thus t$_{sw3}$ $>$ t$_{sw4}$. However, if one desired to switch disk 1 (K$_{z1}$) and disk 4 (K$_{z4}$), the application of magnetic field will switch the disks with intermediate values of K$_z$, such as K$_{z2}$ and K$_{z3}$.\\
\indent Final states with three red digits can be obtained in three steps, following similar procedure, as in the case of two red digits.\\
\indent Full switching of all disks is trivial and can be obtained using a frequency equal to that of the gyrotropic mode.\footnote{The same methodology can be used for the case of having the disks located in the form of nanopillars.}\\
\indent Switching from negative polarity p = -1 to positive polarity p = +1  is also possible using the  azimuthal modes, but changing the sense of rotation of these modes. For p = +1, we have m = +1 (CCW) and m = -1 (CW), and for p = -1 we have m = +1 (CW) and m = -1 (CCW) \cite{Kammerer2011}.\\
\section{Conclusion}
In this work, we initially studied the influence of PUA on azimuthal modes in a matrix of disks with magnetic vortex configuration, using micromagnetic simulations. Our results show that azimuthal mode frequencies decrease with increase of PUA, and modified the intensity of the magnetic field necessary to switching the vortex core. \\
\indent Based on this initial study, we then demonstrated that the azimuthal modes can be used to selective switching in arrays of disks, therefore obtaining several different final state configurations using  a single array of disks. This shows the great advantage of using the azimuthal modes in comparison to the use of the gyrotropic mode.\\ 
\indent This work also shows that when the intrinsic variable K$_z$ is considered, the universality of the value of the critical velocity is broken, even for the gyrotropic mode.\\
\indent  Our proposal addresses a subject not very studied, the influence of the PUA in the magnetic vortex dynamics, allowing to write information in an array of disks through selective switching of vortex cores. This simple system can be expanded to larger arrays. 
\section*{ACKNOWLEDGMENTS}
The authors would like to thank the support of the Brazilian agencies CNPq and FAPERJ.

\section*{References}

\bibliography{referencias}

\end{document}